\documentclass[]{aastex631}

\usepackage{soul}
\usepackage{multirow}
\usepackage{hyperref}
\usepackage{amsmath}
\begin{document}

%\linenumbers

\title{Identifying Ring Galaxies in DESI Legacy Imaging Surveys Using Machine Learning Methods
%\footnote{Released on March, 1st, 2021}
}

\author {Aina Zhang}
\affiliation{ School of Mechanical, Electrical \& Information Engineering, Shandong University, Weihai, 264209, Shandong, People's Republic of China}

\author{Xiaoming Kong}
\affiliation{ School of Mechanical, Electrical \& Information Engineering, Shandong University, Weihai, 264209,  Shandong, People's Republic of China}
\affiliation{ Shandong Key Laboratory of lntelligent Electronic Packaging Testing and Application, Shandong University, Weihai, 264209, Shandong, People's Republic of China}
\email{xmkong@sdu.edu.cn}

\author {Bowen Liu}
\affiliation{ School of Mechanical, Electrical \& Information Engineering, Shandong University, Weihai, 264209, Shandong, People's Republic of China}

\author{Nan Li}
\affiliation{National Astronomical Observatories, Chinese Academy of Sciences, Beijing 100101, China}
\affiliation{School of Astronomy and Space Science, University of Chinese Academy of Science, Beĳing 100049, China}

\author{Yude Bu}
\affiliation{ School of Mathematics and Statistics, Shandong University, Weihai, 264209, Shandong, People's Republic of China}

\author{Zhenping Yi}
\affiliation{ School of Mechanical, Electrical \& Information Engineering, Shandong University, Weihai, 264209,  Shandong, People's Republic of China}
\affiliation{ Shandong Key Laboratory of lntelligent Electronic Packaging Testing and Application, Shandong University, Weihai, 264209, Shandong, People's Republic of China}
\author{Meng Liu}
\affiliation{ School of Mechanical, Electrical \& Information Engineering, Shandong University, Weihai, 264209, Shandong, People's Republic of China}
\affiliation{ Shandong Key Laboratory of lntelligent Electronic Packaging Testing and Application, Shandong University, Weihai, 264209, Shandong, People's Republic of China}



\begin{abstract}
The formation and evolution of ring structures in galaxies are crucial for understanding the nature and distribution of dark matter, galactic interactions, and the internal secular evolution of galaxies. However, the limited number of existing ring galaxy catalogs has constrained deeper exploration in this field. To address this gap, we introduce a two-stage binary classification model based on the Swin Transformer architecture to identify ring galaxies from the DESI Legacy Imaging Surveys. This model first selects potential candidates and then refines them in a second stage to improve classification accuracy.
During model training, we investigated the impact of imbalanced datasets on the performance of the two-stage model. We experimented with various model combinations applied to the datasets of the DESI Legacy Imaging Surveys DR9, processing a total of 573,668 images with redshifts ranging from $z\_spec$ = 0.01-0.20 and $mag_r$ \textless 17.5. After applying the two-stage filtering and conducting visual inspections, the overall $Precision$ of the models exceeded 64.87\%, successfully identifying a total of 8052 newly discovered ring galaxies. With our catalog, the forthcoming spectroscopic data from DESI will facilitate a more comprehensive investigation into the formation and evolution of ring galaxies.

\end{abstract}

%% The AAS Journals now uses Unified Astronomy Thesaurus concepts:A


%\titleformat{\section}{\bfseries\centering}{\thesection}{1em}{}

\section{Introduction} \label{sec:intro}

Ring galaxies are a unique class of galaxies characterized by a sparsely populated or empty central region encircled by a prominent outer ring. 
This ring structure is typically elliptical or circular in shape, and comprises a dense distribution of stars, gas and dust. 
The formation mechanisms of ring galaxies can be summarized in two main ways. First, the gravitational torque effect of the galaxy bars leads to the accumulation of gas in certain resonance regions, forming resonance rings \citep{1996FCPh...17...95B}. This phenomenon originates from dynamical processes within galaxies, where material is driven into specific orbits, resulting in the formation of ring structures. Second, galaxy minor interactions may also trigger the formation of a variety of ring structures, including accretion rings, polar rings, and collisional rings. 
The formation of these rings depends on the mass, gas content of the satellite galaxies and the characteristics of their orbits around large spiral galaxies\citep{1976ApJ...209..382L,1987ApJ...320..454S,2008MNRAS.389.1275D,2022MNRAS.516.3692S}. 
These interactions can lead to the redistribution of material, either through the direct collision of galaxies or through tidal forces, resulting in the striking ring structures observed in these galaxies.

The study of ring galaxies is of significant scientific importance. 
These galaxies can offer extensive insights into the evolutionary and dynamical characteristics of galaxies  \citep{1979ApJ...227..714K, 1996FCPh...17...95B}.
By examining ring galaxies, we can deepen our understanding of the physical processes and structural changes that occur in galaxies at various evolutionary stages. Furthermore, since the formation of ring galaxies is closely linked to interactions between galaxies, studying them can shed light on the mechanisms and impacts of these galactic interactions.

The early discovery of ring galaxies primarily relied on visual inspection by professional astronomers \citep[e.g.][]{1995ApJS...98..739B, 2009ApJS..181..572M, 2010ApJS..186..427N, 2015hsa8.conf..372G}. 
With technological advancements, citizen science projects have increasingly played a significant role in the identification of ring galaxies. 
Although these projects still rely on visual recognition by the human eye, large-scale public participation has greatly enhanced the efficiency and scope of ring galaxy identification. 
For example, \citet {2016MNRAS.461.3663H} classified galaxy morphologies through the Galaxy Zoo2 citizen science project and successfully identified a number of ring galaxies. 
Similarly, \citet {2023PASJ...75..986T} also identified a group of ring galaxies in the GALAXY CRUISE citizen science project.

The accuracy of these catalogs is high because the human brain excels at classifying galaxy morphologies. 
However, since visual inspection requires a large amount of manpower, even with numerous volunteers, it remains difficult to thoroughly analyze all the images collected by modern digital astronomical surveys. To address this challenge, scientists have begun to explore computational analysis to identify ring galaxies. 
\citet {2017ApJS..231....2T} applied a flood-fill algorithm to the PanSTARRS DR1 \citep{2017AAS...22922303C} dataset, successfully identifying 185 ring galaxy candidates. 
Later, \citet {2020MNRAS.491.3767S} further used the algorithm to find 443 ring galaxy candidates in the SDSS DR14 \citep{2018ApJS..235...42A} dataset. 
Although the flood-fill algorithm improved identification efficiency, its classification accuracy was still limited by factors such as off-centered objects, image noise, and artifacts, resulting in an accuracy that has not yet reached the desired level.

With the advancement of machine learning techniques, machine learning methods have achieved remarkable results in the identification and study of special  astronomical objects. As a result, astronomers have begun to apply these techniques to the discovery and classification of ring galaxies. 
For example, \citet {Krishnakumar_2024} first trained an Inception-ResNet-V2\citep{szegedy2017inception} model on 100,000 simulated datasets and then transferred the trained model to real datasets containing ring galaxies for further training and fine-tuning. Finally, the model successfully identified 1967 ring galaxies from a dataset of 960,000 galaxy samples obtained from the \citet{2020ApJS..251...28G} catalog, achieving an identification $Precision$ of 58.9\%. 
\citet {2024PASJ...76..191S} trained a deep learning classifier and applied it to the extensive galaxy sample from the Hyper Suprime-Cam Subaru Strategic Program, ultimately generating a catalog of 33,993 galaxies identified by the model as ring candidates.
Furthermore, \citet{2024arXiv240404484A} classified galaxies in the SDSS DR18 \citep{2023ApJS..267...44A} dataset using an improved AlexNet\citep{krizhevsky2012imagenet} algorithm, ultimately identifying 4855 candidates with ring-like structures, with a prediction confidence exceeding 90 percent. At the same time, they utilized the trained model to identify 2087 candidates with both bar and ring structures from the catalog provided by \citet{2018MNRAS.477..894A}.
These studies demonstrate the potential and practical effectiveness of machine learning technologies in the identification of ring galaxies.

In this study, we imploy a machine learning approach to identify ring galaxies in DESI Legacy Imaging Surveys. The model comprises two binary classifiers. The first classifier is designed to distinguish ring galaxies from all other non-ring galaxies, serving as an initial filter to identify ring galaxy candidates from a large image dataset. Although the model shows high $Precision$ on the test set, there are still more misclassifications when applied to massive data, especially misidentifying spiral and barred spiral galaxies as ring galaxies. To solve this problem, we further trained a second binary classifier for distinguishing ring galaxies from spiral and barred spiral galaxies, thus purifying the initially screened candidates. Finally, we applied the model to the datasets of DESI Legacy Imaging Surveys, conducted visual inspections after identifying ring galaxy candidates, and compiled a new catalog of ring galaxies.

The organization of this paper is as follows: in Section \ref{sec:data}, we provide a detailed description of the dataset used for training, validation, and testing, along with an overview of the data augmentation techniques applied. In Section \ref{sec:methods}, we briefly describe the model architecture used in the study. Section \ref{sec:experiment} presents the specific results and performance of the two binary classification models, along with a comparative evaluation of the algorithms. Section \ref{sec:results} reports the findings from the search for ring galaxies across more than 500,000 galaxy images and offers a comparative analysis of the distribution patterns of ring and non-ring galaxies. A brief conclusion is provided in Section \ref{sec:conclusion}.

\section{Data Set} \label{sec:data}

\subsection{DESI Legacy Imaging Surveys} \label{subsec:desi}

The DESI Legacy Imaging Surveys \citep{2019AJ....157..168D} is a comprehensive astronomical survey resulting from the collaboration of three major public projects: the DECam Legacy Survey (DECaLS), the Beijing-Arizona Sky Survey (BASS), and the Mayall z-band Legacy Survey (MzLS). The DESI Legacy Imaging Surveys cover approximately 14,000 square degrees of the extragalactic sky visible from the Northern Hemisphere, utilizing three key optical bands (g, r, and z). In terms of observational layout, the survey footprint is clearly divided into two regions: the northern and the southern regions. The northern region (Declination \textgreater 32.375) is observed by a combination of BASS (g, r-band) and MzLS (z-band) to ensure data comprehensiveness and accuracy, while the southern region (Declination \textless 32.375) is primarily covered by DECaLS(g, r, z-band). All images used in this study are sourced from the DESI Legacy Imaging Surveys Data Release 9 (hereafter DR9).

\subsection{Data Sample Construction} \label{subsec:construction}

This study aims to identify ring galaxies in the DR9 dataset by developing a classifier tailored for ring galaxy recognition. To train and evaluate the classifier, we constructed two datasets: a positive set of ring galaxies and a negative set of non-ring galaxies.

For the positive sample, we cross-matched the DR9 catalog with 2598 ring galaxies from \citet{2010ApJS..186..427N}, 185 ring galaxies from \citet{2017ApJS..231....2T}, 443 ring galaxies from \citet{2020MNRAS.491.3767S}, and 1151 ring galaxies from \citet{2022arXiv221011428K}, resulting in a total of 4377 ring galaxies. Additionally, we selected 3657 ring galaxies from Galaxy Zoo 2 \citep {2016MNRAS.461.3663H} that meet the criteria `` t08\_odd\_feature\_a19\_ring\_debiased $\geq 0.9999$ $\cap$ t08\_odd\_feature\_a19\_ring\_count $\textgreater 1$ '' and 853 ring galaxies from the GALAXY CRUISE Season 1 Data Release \citep {2023PASJ...75..986T} that meet the condition ``  P(interaction)  $\textgreater 0.5$  $\cap$  P(ring) $\textgreater$ [P(fan), P(tail), P(distorted)] ''. After identifying these ring galaxies, we extracted images centered on each target's right ascension (RA) and declination (Dec) from the DR9 dataset, obtaining g, r, and z-band images with a pixel scale of 0.262 $\prime$$\prime$/pixel and dimensions of 256×256 pixels. In total, we acquired 8887 images of ring galaxies.

\begin{figure}[ht!]
	\plotone{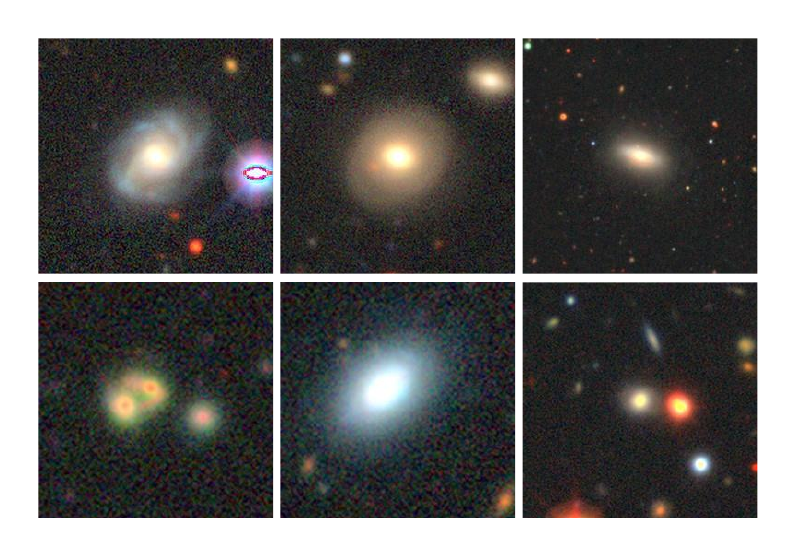}
	\caption{Examples of samples screened based on visual inspection.  \label{fig:deleted-pictures}}
\end{figure}

To ensure the quality of the galaxy images in the positive sample and eliminate particularly challenging cases,
%\sout{To ensure the quality and representativeness of the galaxy }
%\sout{images in the positive sample,} 
we visually inspected each image and excluded those
containing multiple celestial sources (i.e., images with multiple distinct objects rather than a single ring galaxy) and those with ring features that were either not prominent or difficult to clearly identify. Through this visual screening process, we removed 4774 images that did not meet the criteria. Figure \ref{fig:deleted-pictures} shows examples of some of the excluded samples. Ultimately, the study included 4113 verified ring galaxy samples. Table \ref{tab:description} provides a detailed account of the number of ring galaxies contributed by each referenced study in this research.

\begin{deluxetable*}{c c }
	\tablenum{1}
	\setlength{\tabcolsep}{6mm}
	\tablecaption{Description of the Ring Galaxies Samples Used in Our Study}\label{tab:description}
	\tablewidth{0pt}
	\tablehead{
		\colhead{Source} & \colhead{Number of used ring galaxies} 		
	}
	\startdata
	\citet{2010ApJS..186..427N} & 1087 \\
	\citet{2016MNRAS.461.3663H} & 1726 \\
	\citet{2017ApJS..231....2T}	& 71 \\	
	\citet{2020MNRAS.491.3767S} & 252 \\
	\citet{2022arXiv221011428K} & 774 \\
	\citet{2023PASJ...75..986T} & 203 \\
	 &$\sum=4113$ \\
	\enddata
\end{deluxetable*}

In constructing the negative sample set, we used the galaxy classification system from the Galaxy Zoo 2 project \citep{2016MNRAS.461.3663H}. Galaxy Zoo 2 is a citizen science astronomy project designed to classify galaxy images from SDSS DR7 \citep{2009ApJS..182..543A} with the help of volunteers. Volunteers identify the shape and structure of galaxies based on a series of questions that form a decision tree. Each volunteer answers only a subset of these questions, and the final classification is determined by aggregating all volunteers' responses to calculate a weighted voting score, which estimates the confidence of the classification. Using this classification system, we selected non-ring galaxies from the Galaxy Zoo 2 database with ``t08\_odd\_feature\_a19\_ring\_count \textless 1.'' These galaxies were then cross-matched with the DR9 dataset to obtain common galaxy records. Then, galaxies with $z\_spec = 0.01$-$0.20$ and $mag_r$ \textless 17.5  were further filtered out. After constructing the catalog of non-ring galaxies, we extracted image data for these galaxies from DR9 with a uniform size of 256 × 256 pixels.

\subsection{Data Augmentation} \label{subsec:augmentation}

Data augmentation aims to create a more diverse and enriched sample set by transforming and expanding the existing data, thereby improving the training effectiveness of machine learning models. This technique is thoroughly demonstrated and applied in \citet{2018MNRAS.476.5365S}. In our study, due to the relatively limited number of ring galaxy samples, we applied the following data augmentation techniques to expand the dataset and enhance model performance:

(1) Rotation: Images were rotated at fixed angles of 90°, 180°, and 270°, as well as random rotations between 10° and 350°, generating a total of 4 augmented images.

(2) Translation: Images were horizontally shifted by 5 to 25 pixels in either direction (left or right), resulting in 2 translated images.

(3) Horizontal/Vertical Mirroring: Images were flipped along the horizontal and vertical axes, producing 2 mirrored images.

(4) Center Cropping and Scaling: The original 256×256 pixel images were center-cropped at 80\% of their original size, then scaled back to 256×256 pixels.

(5) Color Jitter: Random adjustments were made to the brightness, contrast, saturation, and hue of the images, with an adjustment range of 0.2. For instance, brightness was randomly altered between 80\% (1-0.2) and 120\% (1+0.2) of the original value.

\begin{figure}[ht!]
	%\plotone{augmentation.pdf}
	 \includegraphics[width=1.0\columnwidth]{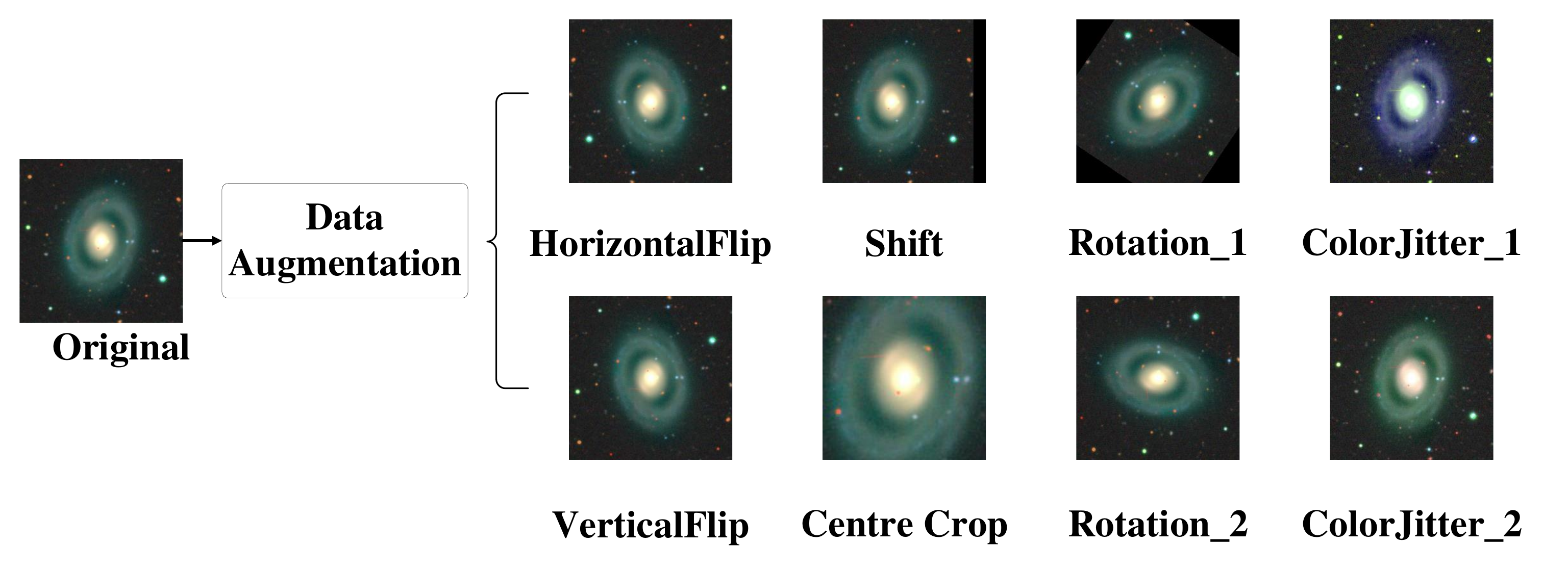}
	 %\centerline{\includegraphics[width=0.7\textwidth]{augmentation.pdf}}
	\caption{The original image and the results of 8 samples after applying data augmentation used in this work.  \label{fig:augmentation}}
\end{figure}

During the experiment, we randomly applied the above augmentation techniques to each original image to increase the diversity and quantity of positive samples in the training set.  Figure \ref{fig:augmentation} illustrates the effects of these augmentation techniques on a sample ring galaxy. These data augmentation strategies allowed us to successfully generate and store a more diverse set of images, thereby expanding the sample size of the dataset. Before feeding all images into the model, we resized them to 224×224 pixels and performed normalization.

\section{Methods} \label{sec:methods}

\subsection{Swin Transformer} \label{subsec:swin}

The Swin Transformer \citep{liu2021swin}, proposed by Microsoft Research Asia, is a deep learning model based on the Transformer architecture, specifically designed for image processing tasks. It has demonstrated outstanding performance in the field of computer vision, particularly in image classification tasks. The key innovation of this model lies in the introduction of a windowed local self-attention mechanism, which restricts the computation of self-attention to non-overlapping windows, and a shifted window scheme that enables cross-window information exchange. Additionally, the model adopts a hierarchical structure to facilitate multi-scale feature learning. These design choices significantly enhance the model’s computational efficiency and its ability to capture spatial information at different scales, addressing the computational complexity issues associated with the original Transformer in vision tasks, while providing more robust and flexible feature learning capabilities.

\begin{figure*}[ht!]
	%\plotone{swin.pdf}
	\includegraphics[width=\textwidth]{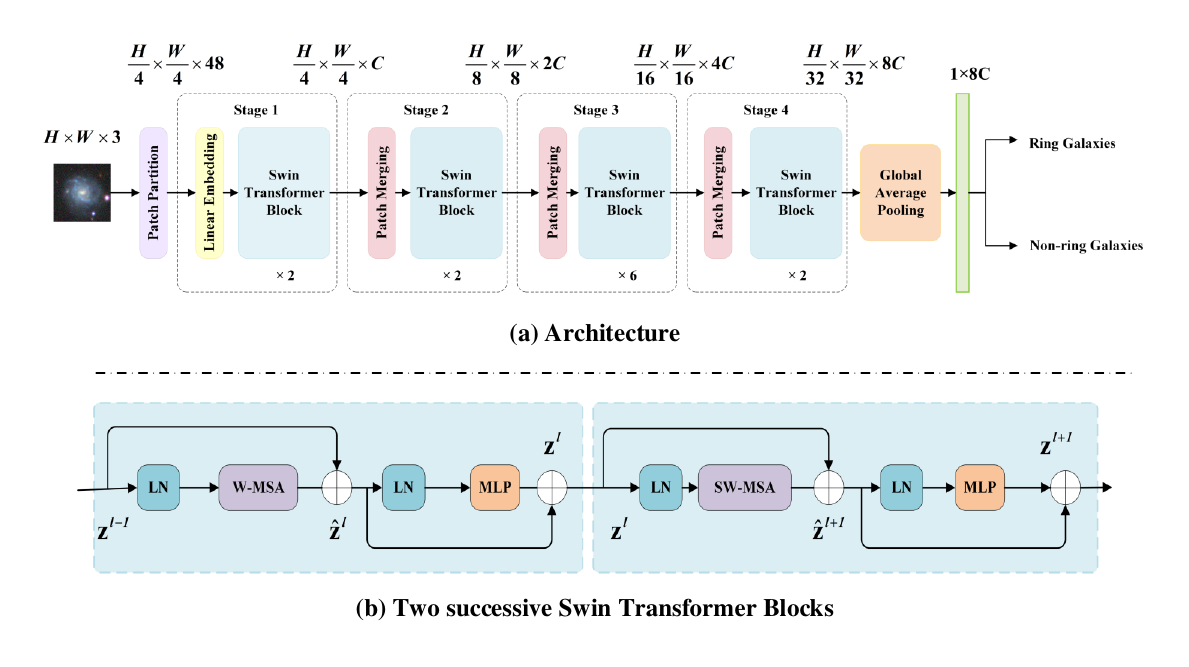}
	\caption{(a) Swin Transformer architecture; (b) Swin Transformer Blocks. \label{fig:swin}}
\end{figure*}

Figure \ref{fig:swin} provides an overview of the architecture of the Swin Transformer, specifically illustrating the Tiny version (Swin-T) used in this study. First, the input galaxy images are processed by the Patch Partition layer, which divides the images into multiple patches for subsequent processing. These patch sequences then pass through four stages, progressively constructing multi-scale feature maps. In Stage 1, the Linear Embedding layer projects each patch to a higher dimension (C=96), transforming the entire image into a feature sequence. Subsequently, 2 Swin Transformer blocks perform initial feature extraction on the sequence. In Stage 2, the Patch Merging layer reduces the number of patches and increases the number of channels, followed by 2 more Swin Transformer blocks that extract deeper features. Stage 3 mirrors the process of Stage 2, where the Patch Merging layer again reduces the patch count and increases the number of channels, and 6 Swin Transformer blocks further enhance feature extraction. In Stage 4, the Patch Merging layer reduces the number of patches for the third time, increasing the channel count to its maximum value. This is followed by 2 final Swin Transformer blocks to integrate the extracted features and ensure all critical information is captured. To classify the galaxy images, the resulting feature sequence is transformed into a vector via a global average pooling layer, simplifying it into key feature representations. Finally, a softmax layer maps these features to a probability distribution across the classes, completing the galaxy image classification process.

The Swin Transformer block, as illustrated in Figure \ref{fig:swin}(b), employs a shifted window-based Multi-Head Self-Attention (MSA) mechanism. Each block consists of two key components: an MSA module and a Multi-Layer Perceptron (MLP) with a GELU activation function. LayerNorm (LN) is applied before both the MSA and MLP layers to stabilize the training process. Additionally, residual connections are incorporated after each module to enhance the model's deep learning capabilities. This architecture includes two types of MSA: Window-based Multi-Head Self-Attention (W-MSA) and Shifted Window-based Multi-Head Self-Attention (SW-MSA). W-MSA focuses on computing self-attention within fixed windows, allowing the model to capture local details by focusing on smaller regions of the image. In contrast, SW-MSA introduces a shifted window mechanism, which offsets the window positions during attention computation. This shift improves the model's ability to capture long-range dependencies, and helps to deal with the effective transfer of global information.

\section{Experiment} \label{sec:experiment}

\subsection{Performance Metric} \label{subsec:performance}

The objective of our task is to accurately classify ring galaxies and non-ring galaxies. In evaluating the performance of the classification model, we employed three metrics: $Precision$, $Recall$, and $F1\_score$, which are defined as follows:

\begin{eqnarray}
	Precision=\frac{TP}{TP+FP},\\
	Recall=\frac{TP}{TP+FN},\\
	F1\_score=\frac{2\times Precision\times Recall}{Precision+Recall},
\end{eqnarray}

Here, True Positives (TP) represent the number of actual ring galaxies correctly identified by the model. False Positives (FP) denote the number of non-ring galaxies that the model incorrectly classified as ring galaxies. False Negatives (FN) refer to the actual ring galaxies that the model mistakenly classified as non-ring galaxies. $Precision$ is the proportion of true ring galaxies among the samples predicted as ring galaxies by the model. $Recall$ is the proportion of actual ring galaxies correctly classified by the model out of the total number of ring galaxies. The $F1\_score$, which is the harmonic mean of $Precision$ and $Recall$, provides a balanced evaluation of the classifier's performance. By employing these metrics, we can comprehensively assess the model’s ability to distinguish between ring and non-ring galaxies, offering valuable insights into its classification effectiveness.

\subsection{Classification Construction} \label{subsec:classification}

In this section, we introduce the development of a two-stage model for ring galaxy classification. In the first stage, we constructed a binary classifier based on the Swin Transformer to perform an initial screening of potential ring galaxies from a large sample of galaxies. In the second stage, we developed a new binary classifier to further filter out spiral and barred spiral galaxies that may have been misclassified in the first stage. By adopting this two-stage approach, we aim to reduce false positives, thereby enhancing the reliability and efficiency of the classification system when applied to the DESI Legacy Imaging Surveys DR9 dataset.

\subsubsection{Stage 1}

In constructing training datasets for machine learning algorithms, issues with imbalanced sample distribution frequently arise, particularly in the search for rare celestial objects, where the number of positive samples is typically scarce. Properly constructing the training set is crucial for optimizing model performance. To this end, we conducted experiments to evaluate the impact of the ratio of positive to negative training samples on the model's performance, while ensuring both sample quantity and diversity.

In the first stage, we selected a sample set comprising 4113 ring galaxy images and 35,000 non-ring galaxy images, as described in detail in Section \ref{subsec:construction}. These samples were randomly split into training, validation, and test sets with a ratio of 8:1:1. Following the data augmentation strategies outlined in Section \ref{subsec:augmentation}, we augmented the positive samples in the training set, increasing their numbers to threefold and eightfold the original size.
Using these three different configurations of the training set, we independently trained the Swin Transformer model. 
The model trained on the original training set was labeled Swin T1$_{1\text{-}8}$, while the models trained on the threefold and eightfold augmented sets were named Swin T1$_{3\text{-}8}$ and Swin T1$_{8\text{-}8}$, respectively.

\begin{figure}[ht!]
	%\plotone {redshift.pdf}
	\centerline{\includegraphics[width=0.6\textwidth]{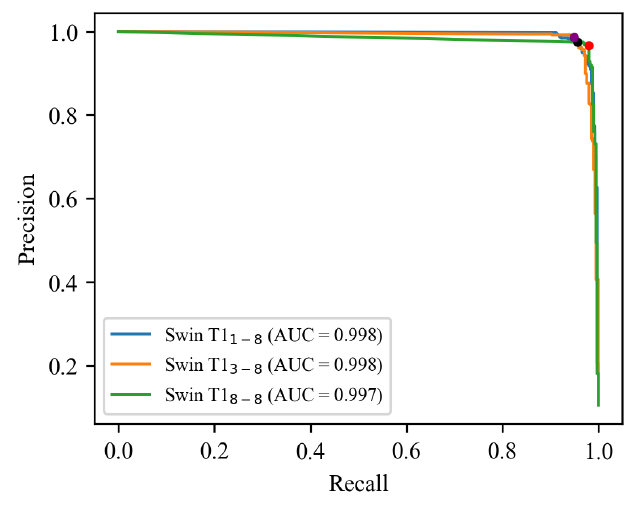}}
	\caption{PR curve for each model. The AUC values are identical across all three models, and the black, purple, and red points in the graph indicate the maximum $F1\_score$ achieved by Swin T1$_{1\text{-}8}$, Swin T1$_{3\text{-}8}$, and Swin T1$_{8\text{-}8}$, which are 96.57\%, 96.78\%, and 97.35\%, respectively.  \label{fig:pr2}}
\end{figure}

We evaluated the performance of these classifiers using precision-recall (PR) curves (see Figure \ref{fig:pr2}). The experimental results demonstrate that the three models have very similar overall performance, with their AUC values differing by only 0.001. The maximum $F1\_scores$ indicate that both Swin T1$_{3\text{-}8}$ and Swin T1$_{8\text{-}8}$ outperform Swin T1$_{1\text{-}8}$, with Swin T1$_{8\text{-}8}$ achieving the highest maximum $F1\_score$. Table  \ref{tab:stage1} summarizes the $Precision$, $Recall$, and $F1\_score$ for each model at a classification threshold of 0.5. Notably, Swin T1$_{3\text{-}8}$ achieves the highest $Precision$, indicating a lower contamination rate (fewer false positives), while Swin T1$_{8\text{-}8}$ achieves the highest $Recall$, suggesting better sample completeness but with more false positives. These differences are particularly important when applying the models to large-scale real-world data for ring galaxy identification, where the trade-off between contamination and completeness must be carefully considered. Therefore, we selected Swin T1$_{3\text{-}8}$ and Swin T1$_{8\text{-}8}$ for further analysis in order to investigate their respective performance differences on the application dataset.

\begin{deluxetable*}{c c c c c}
	\tablenum{2}
	\setlength{\tabcolsep}{6mm}
	\tablecaption{Results of the First Binary Classification Model}\label{tab:stage1}
	\tablewidth{0pt}
	\tablehead{
		\colhead{Models} & \colhead{$Precision$} & \colhead{$Recall$} & \colhead{$F1\_score$} &\colhead{Ring Galaxies Identified in DR9}	
	}
	\startdata
	Swin T1$_{1\text{-}8}$ & 96.58\% & 95.87\% & 96.22\% & \nodata \\
	Swin T1$_{3\text{-}8}$	& 98.73\% & 94.42\% & 96.53\% & 49,264 \\
	Swin T1$_{8\text{-}8}$	& 96.42\% & 98.06\% & 97.23\% & 72,022 \\	
	\enddata
\end{deluxetable*}

\subsubsection{Stage 2}

In the second stage of the classification task, we first analyzed the results from the first stage and found that most of the misclassified samples were concentrated in spiral and barred spiral galaxies. Therefore, we constructed a classifier to further eliminate these contaminating samples. During the selection of the negative sample dataset, our experimental results indicate that when the numbers of spiral and barred spiral galaxies are balanced within the negative samples, the performance of the trained model is significantly enhanced. To achieve this, we employed a random sampling method to ensure, to the greatest extent possible, that a 1:1 ratio is maintained between these two types of galaxy samples.

At this stage, the negative samples consist only of spiral and barred spiral galaxies, resulting in significantly lower sample complexity compared to the first stage. We controlled the size of the negative sample set and investigated the impact of negative sample quantity and class balance on model performance.
We kept the number of positive samples constant at 4113 ring galaxy images and adjusted the negative sample quantities to be 2, 3, and 5 times that of the positive samples, creating three datasets with positive-to-negative ratios of 1:2, 1:3, and 1:5, respectively, for model training. 
Additionally, we performed upsampling on the positive samples in each of these training sets to match the number of negative samples for comparative experiments. We trained a total of six models through a series of experiments. The specific model names (e.g., Swin T2$_{1\text{-}2}$, Swin T2$_{1\text{-}3}$, Swin T2$_{1\text{-}5}$ represent models with original ratios, while Swin T2$_{2\text{-}2}$, Swin T2$_{3\text{-}3}$, Swin T2$_{5\text{-}5}$ represent models with augmented ratios) reflect the characteristics of their training sets, facilitating subsequent analysis and comparison.

Table \ref{tab:stage2} presents a comprehensive evaluation of six models on both the test set and in practical applications. As shown, the $F1\_scores$ of all models vary by less than 2\%, demonstrating stable performance across different configurations.
Additionally, models trained on balanced datasets with data augmentation generally performed better on the test set compared to those trained on imbalanced datasets. Overall, Swin T2$_{2\text{-}2}$ and Swin T2$_{3\text{-}3}$ exhibited superior performance. Based on the above analysis, we retained the Swin T2$_{2\text{-}2}$, Swin T2$_{3\text{-}3}$, and Swin T2$_{5\text{-}5}$ models, which were trained on balanced datasets, to compare their performance on the application dataset.

\begin{deluxetable*}{c c c c c}
	\tablenum{3}
	\setlength{\tabcolsep}{6mm}
	\tablecaption{Results of the Second Binary Classification Model}\label{tab:stage2}
	\tablewidth{0pt}
	\tablehead{
		\colhead{Models} & \colhead{$Precision$} & \colhead{$Recall$} & \colhead{$F1\_score$} &\colhead{$Precision$ at DR9}	
	}
	\startdata
	Swin T2$_{1\text{-}2}$	& 94.51\% & 91.99\% & 93.23\% & \nodata \\
	Swin T2$_{2\text{-}2}$	& 97.67\% & 91.74\% & 94.61\% &  10,292/15,785=65.20\% \\
	Swin T2$_{1\text{-}3}$	& 96.17\% & 91.50\% & 93.78\% & \nodata \\
	Swin T2$_{3\text{-}3}$	& 96.45\% & 92.23\% & 94.29\% & 10,540/16,320=64.58\% \\
	Swin T2$_{1\text{-}5}$	& 97.61\% & 89.08\% & 93.15\% & \nodata \\
	Swin T2$_{5\text{-}5}$	& 96.40\% & 91.02\% & 93.63\% & 7551/11,374=66.39\% \\	
	\enddata
\end{deluxetable*}

%\subsection{Comparison with Other Methods} \label{subsec:comparison}
\subsection{Model Performance} \label{subsec:comparison}

\begin{figure*}[ht!]
	%\plotone {redshift.pdf}
	%\includegraphics[width=0.8\columnwidth]{redshift.pdf}
	%\centerline{\includegraphics[width=\textwidth]{pr4.pdf}}
	\includegraphics[width=\textwidth]{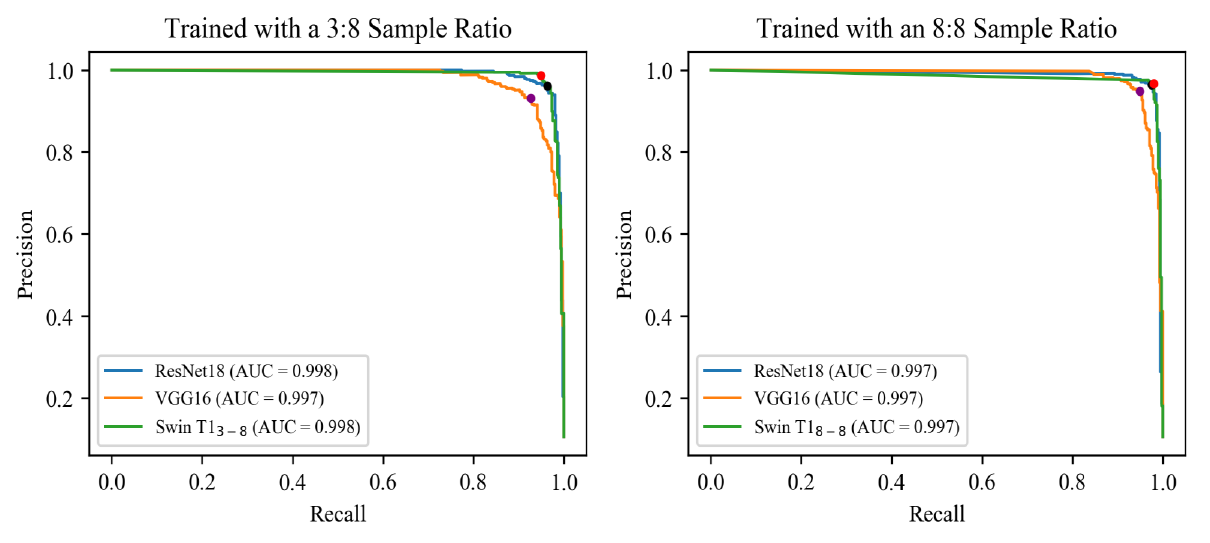}
	\caption{PR curve for each model. The AUC values are identical across all three models, and the black, purple, and red points in the graph indicate the maximum $F1\_score$ achieved by ResNet18, VGG16, and Swin Transformer respectively. When trained with a positive-to-negative sample ratio of 3:8, the maximum $F1\_scores$ were 96.24\% (ResNet18), 92.94\% (VGG16), and 96.78\% (Swin T1$_{3\text{-}8}$). Increasing the training ratio to 8:8 resulted in improved maximum $F1\_scores$ of 96.62\% (ResNet18), 94.90\% (VGG16), and 97.35\% (Swin T1$_{8\text{-}8}$), respectively.  \label{fig:pr4}}
\end{figure*}

Before applying the Swin Transformer algorithm to ring galaxy identification in the DESI Legacy Imaging Surveys DR9 dataset, we compared its performance with that of ResNet18 and VGG16. All models were trained on the same datasets as Swin T1$_{3\text{-}8}$ or Swin T1$_{8\text{-}8}$ and evaluated on identical test sets. As shown by the PR curve analysis (Figure \ref{fig:pr4}), both Swin Transformer and ResNet18 outperform VGG16, with the Swin Transformer achieving a slightly higher maximum $F1\_score$. Based on these results, we selected the Swin Transformer algorithm for ring galaxy identification. Table \ref{tab:comparison} summarizes the $Precision$, $Recall$, and $F1\_score$ of each model at a fixed threshold of 0.5.

\begin{deluxetable*}{c c c c c}
	\tablenum{4}
	\setlength{\tabcolsep}{8mm} % 调整列间距
	\tablecaption{Comparison of Classification Performance with Different Training Sample Ratios}
	\label{tab:comparison}
	\tablewidth{0pt}
	\tablehead{
		\colhead{Ratio} & \colhead{Models} & \colhead{$Precision$} & \colhead{$Recall$} & \colhead{$F1\_score$}
	}
	\startdata
	\multirow{3}{*}{3:8} & ResNet18 & 97.22\% & 93.20\% & 95.17\% \\
	& VGG16 & 93.15\% & 92.48\% & 92.81\% \\
	& Swin T1$_{3\text{-}8}$ & 98.73\% & 94.42\% & 96.53\% \\
	\hline
	\multirow{3}{*}{8:8} & ResNet18 & 96.15\% & 97.09\% & 96.62\% \\
	& VGG16 & 95.77\% & 93.45\% & 94.60\%\\
	& Swin T1$_{8\text{-}8}$	& 96.42\% & 98.06\% & 97.23\% \\	
	\enddata
\end{deluxetable*}

Our hardware configuration includes 2 Xeon® Gold 6346 CPUs with a base frequency of 3.10 GHz (16 cores) and 512 GB of DDR4 3200 memory. Additionally, we utilized an NVIDIA Tesla A100 GPU for processing. For the first binary classification model, the average processing time per image is 0.007 seconds, while for the second binary classification model, the average processing time per image is 0.009 seconds.

\section{Results and Discussion} \label{sec:results}
%\section{Results} \label{sec:results}

\subsection{Searching for Ring Galaxies from DESI Legacy Imaging Surveys} \label{subsec:searching}

When using the Swin Transformer algorithm to identify ring galaxies in the DESI Legacy Imaging Surveys DR9 dataset, we first performed data filtering, focusing on galaxy samples with $z\_spec = 0.01$-$0.20$ and $mag_r$ \textless 17.5. We then downloaded images of these selected galaxies, totaling 573,668 images, and proceeded with the search for ring galaxies following the steps outlined below:

Step 1, we applied the binary classifiers Swin T1$_{3\text{-}8}$ and Swin T1$_{8\text{-}8}$ from the first stage to the 573,668 images, identifying 49,264 and 72,022 potential ring galaxy candidates, respectively (see Table \ref{tab:stage1}). This result is consistent with the model's performance on the test set, where Swin T1$_{8\text{-}8}$ exhibited a higher $Recall$ but lower $Precision$ compared to Swin T1$_{3\text{-}8}$. To further validate these findings, we randomly selected 1000 galaxy samples from the difference between the 72,022 and 49,264 candidates for visual inspection, which revealed a $Precision$ of only 9\%. Therefore, we decided to perform further classification on the 49,264 candidates in subsequent applications.

Step 2, we applied the binary classifiers Swin T2$_{2\text{-}2}$, Swin T2$_{3\text{-}3}$, and Swin T2$_{5\text{-}5}$ from the second stage to classify the 49,264 ring galaxy candidates. 
These classifiers identified 15,785, 16,320, and 11,374 positive samples, respectively. Visual inspection (see the fifth column of Table \ref{tab:stage2}) yielded $Precision$ values of 65.20\% ($Precision$ = 10,292/15,785), 64.58\% ($Precision$ = 10,540/16,320), and 66.39\% ($Precision$ = 7551/11,374). These results indicate that, with similar $Precision$ rates, Swin T2$_{2\text{-}2}$ and Swin T2$_{3\text{-}3}$ have comparable $Recall$ rates, which are significantly better than those of Swin T2$_{5\text{-}5}$.

\begin{figure*}[ht!]
	%\plotone{new-rings.pdf}
	\includegraphics[width=\textwidth]{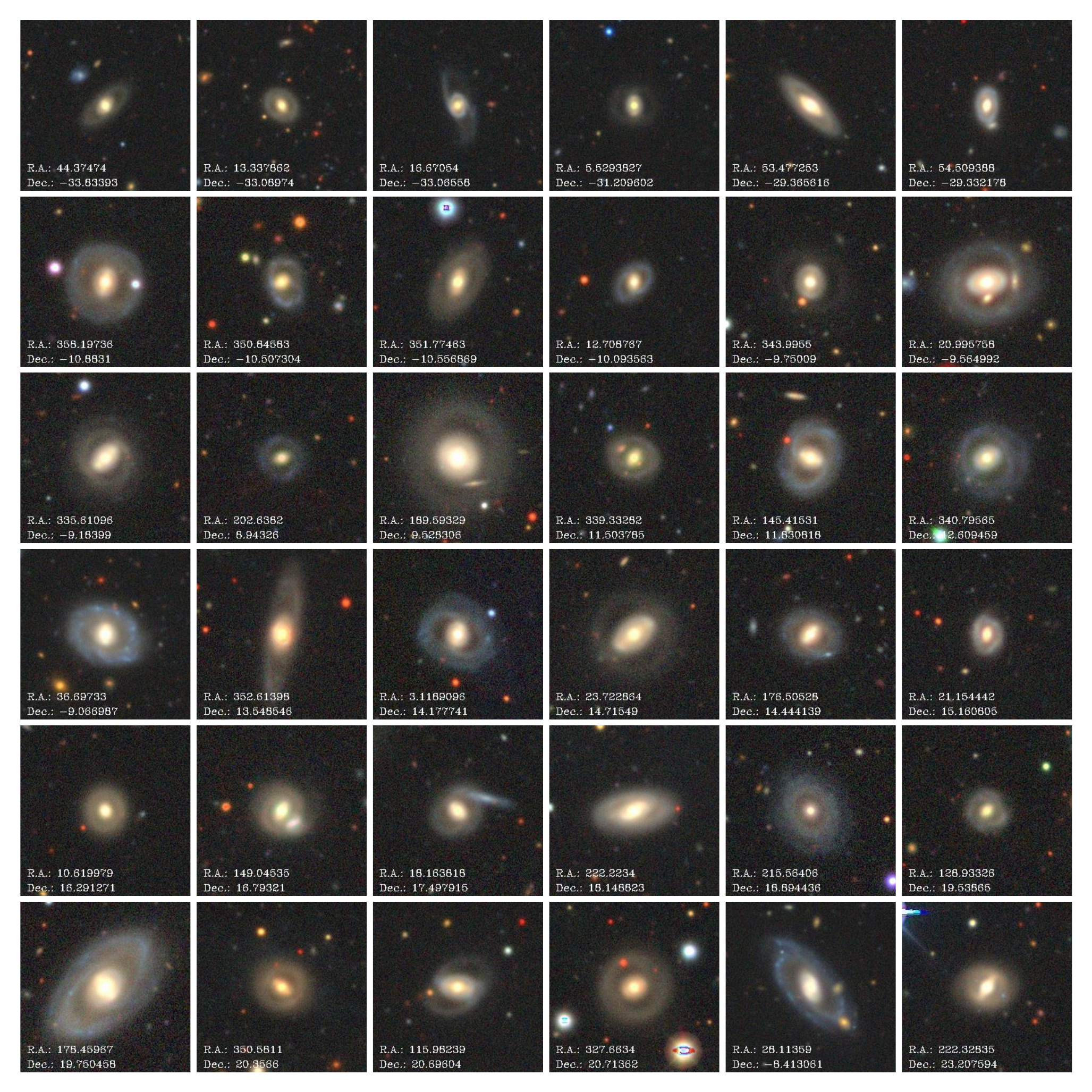}
	\caption{36 galaxies from the catalog of \textcolor{red}{\textbf{8052}} rings.  \label{fig:new-rings}}
\end{figure*}

We combined the candidate sets identified by the three models in the last column of Table \ref{tab:stage2} and removed duplicate entries, resulting in a total of 18,802 unique candidate images. Among these, 12,196 were confirmed as true ring galaxies through systematic visual inspection, corresponding to an overall $Precision$ of64.87\% (12,196/18,802). We further cross-checked these confirmed ring galaxies against our training samples and existing catalogs \citep{2016MNRAS.461.3663H,2024PASJ...76..191S,Krishnakumar_2024,2024arXiv240404484A}, removing 4144 overlapping entries.
Ultimately, we successfully identified 8052 new ring galaxies,
%\sout{Ultimately, we successfully identified and revealed \textcolor{red}{\textbf{8052}} new ring galaxies,}
expanding the current catalog of known ring galaxies. Figure \ref{fig:new-rings} displays images of some of the newly discovered ring galaxy samples from this study. For the convenience of future research, Table \ref{tab:new} provides basic information on a subset of these 8052 newly discovered ring galaxies, including objid, R.A. (right ascension), Dec (declination), z\_spec (the spectroscopic redshift of galaxies \citep{zhou2023desi}), and flux (including the g, r, z bands). The complete dataset is available from the online catalog at DOI: \href{https://doi.org/10.5281/zenodo.15545272}{10.5281/zenodo.15545272}.

\begin{deluxetable*}{r r r r r r r}
	\tablenum{5}
	\setlength{\tabcolsep}{5mm}
	\tablecaption{Information on the 8052 Ring Galaxies Identified in This Study}\label{tab:new}
	\tablewidth{0pt}
	\tablehead{
		 \colhead{objid} & \colhead{R.A.} & \colhead{Dec} & \colhead{z\_spec} & \colhead{flux\_g} & \colhead{flux\_r} & \colhead{flux\_z}	
	}
	\startdata
	1448 & 179.226000 & 32.158806 & 0.010236  & 2157.869000 & 4603.358400 & 8062.846700  \\
	 2398 & 175.415570 & 15.965650 & 0.010757 & 2006.097900 & 4332.462000 & 7971.189500  \\
	2423 & 174.451970 & 22.691252 & 0.011864 & 4115.064500 & 8358.751000 & 14981.679000  \\
	 134 & 175.058060  & 24.697021 & 0.011900 & 5549.910600 & 11949.888000 & 22030.479000  \\
	 3892 & 30.935873 & 14.709282 & 0.012167 & 2941.712400 & 6353.866700 & 11344.872000  
	\enddata
	\tablecomments{Table 5 is published in its entirety in the machine-readable format. A portion is shown here for guidance regarding its form and content.}
\end{deluxetable*}

\subsection{Statistics Properties} \label{subsec:statistics}
%We compared the redshift and color differences of the ring galaxies discovered in this study with those of known ring galaxies and other galaxies. 
This section presents a comparative analysis of the distribution patterns of ring and non-ring galaxies, focusing on redshift and color index ranges.

\begin{figure}[ht!]
	%\plotone {redshift.pdf}
	%\includegraphics[width=0.8\columnwidth]{redshift.pdf}
	\centerline{\includegraphics[width=0.6\textwidth]{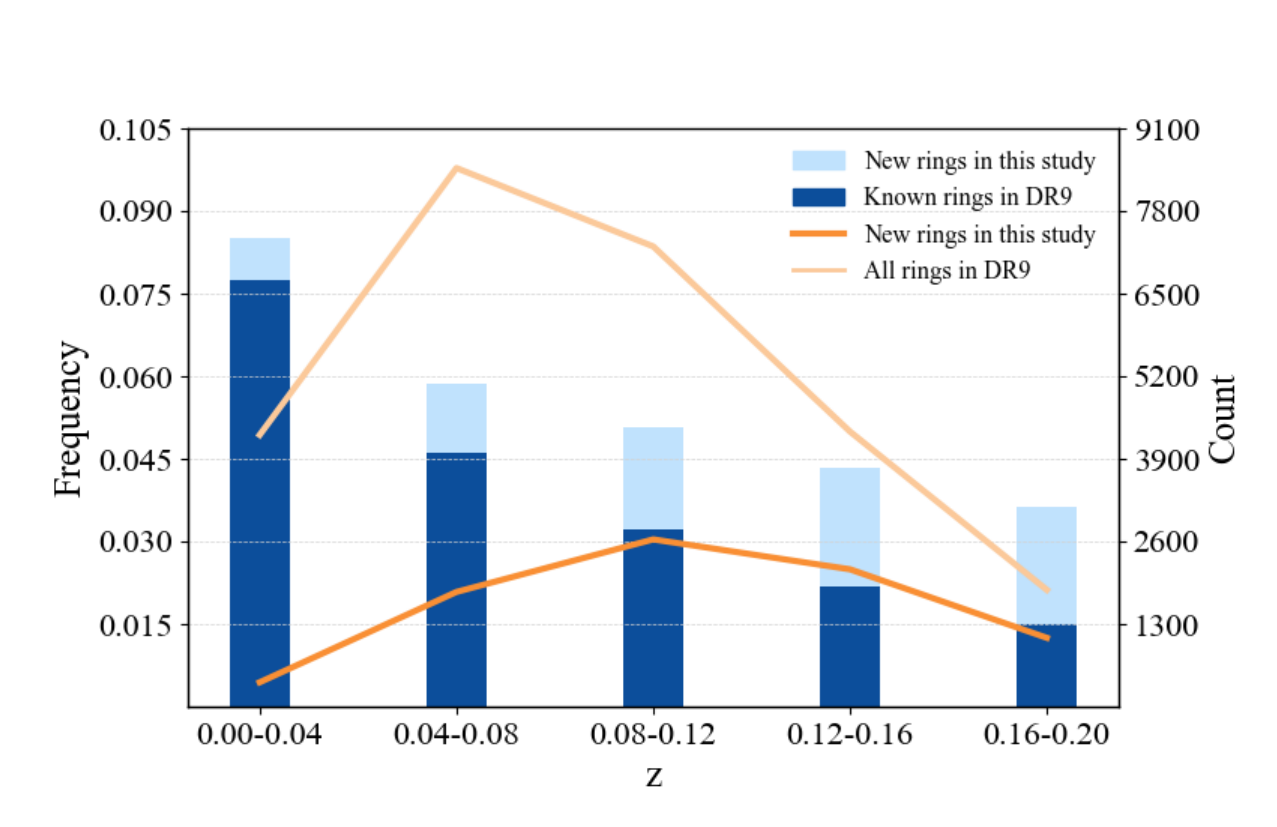}}
	\caption{The number and frequency of newly discovered ring galaxies in this study, as well as those from other known catalogs in DR9 that meet the criteria of $z_{\text{spec}} = 0.01$–$0.20$ and $mag_r < 17.5$. The light blue bars represent the frequency of newly discovered ring galaxies in this study, and the dark blue bars show the frequency of known ring galaxies in DR9. The light orange line denotes the total number of ring galaxies in DR9, and the dark orange line represents the number of newly discovered ring galaxies in this study.  \label{fig:redshift}}
\end{figure}

Figure \ref{fig:redshift} illustrates the redshift distribution of newly discovered and previously cataloged ring galaxies. The known ring galaxies are sourced from multiple catalogs, cross-matched with DR9 galaxies meeting the criteria of $z_{\text{spec}} = 0.01$-$0.20$ and $mag_r < 17.5$. These catalogs include contributions from \citet{2010ApJS..186..427N}, \citet{2016MNRAS.461.3663H}, \citet{2017ApJS..231....2T}, \citet{2020MNRAS.491.3767S}, \citet{2023PASJ...75..986T}, \citet{2024PASJ...76..191S}, \citet{Krishnakumar_2024}, and \citet{2024arXiv240404484A}.  
In Figure \ref{fig:redshift}, the bar chart displays the frequency of ring galaxies within galaxy groups across the same redshift range, while the line chart shows the number of ring galaxies within each redshift interval.

A notable difference appears when comparing these findings to the redshift distribution shown in Figures 6 and 7 of \citet{2020MNRAS.491.3767S}, which reported the lowest proportion of ring galaxies within the lowest redshift range (0.00–0.02). \citet{2020MNRAS.491.3767S} attributed this trend to an excess of objects misclassified as galaxies by the SDSS pipeline that is, in reality, non-extragalactic. Simultaneously, they also pointed out that the small number of detected ring galaxies in their sample in this range limited statistical analysis. In Figure \ref{fig:redshift} of this study, 70.55\% of the known ring galaxies within the redshift range of 0.00–0.04 are sourced from \citet{2024arXiv240404484A}, who identified 4855 ring galaxy candidates (with prediction confidence exceeding 90\%) using an enhanced AlexNet method in SDSS DR18. Among these candidates, 2748 fall within the redshift range of 0.00–0.04, significantly increasing the proportion of ring galaxies within this redshift interval.

Overall, Figure \ref{fig:redshift} indicates a gradual decrease in the frequency of ring galaxies as redshift increases. This trend may be explained by the less detailed morphology of imaged galaxies at higher redshifts, which makes it difficult to identify morphological features, such as the presence of rings.

\begin{figure*}[ht!]
	%\plotone {color.pdf}
	%\includegraphics[width=\textwidth]{color.pdf}
	\includegraphics[width=1.0\textwidth]{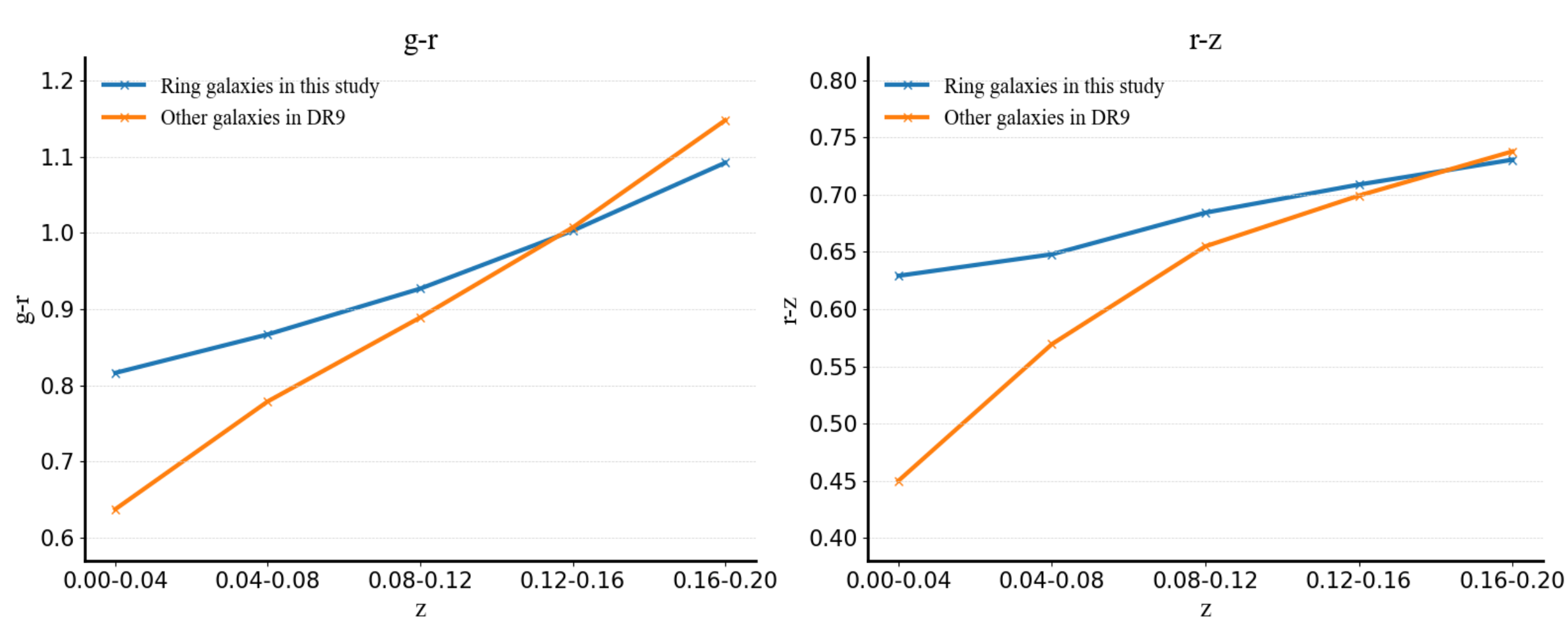}
	\caption{The distribution of color differences between the newly discovered ring galaxies and other galaxies in DR9 that meet the criteria of $z_{\text{spec}} = 0.01$–$0.20$ and $mag_r < 17.5$. The blue line represents the ring galaxies identified in this study, while the orange line denotes the other galaxies in DR9.  \label{fig:color}}
\end{figure*}

Figure \ref{fig:color} shows the color difference distribution between the newly identified ring galaxies in this study and non-ring galaxies in DR9 that meet the criteria of $z_{\text{spec}} = 0.01$-$0.20$ and $mag_r < 17.5$. The results indicate that, for ring galaxies, the g-r and r-z color differences increase with redshift, a trend that parallels that of non-ring galaxies. However, ring galaxies exhibit a less pronounced response to redshift changes in color differences compared to the general galaxy population. This aligns with findings by \citet {2020MNRAS.491.3767S}, suggesting a more homogeneous morphological and evolutionary nature among ring galaxies and their generally lower star formation rates relative to other galaxy types \citep{2024A&A...683A..32F}.

Additionally, we can notice significant differences in two color indices for galaxies within the 
$z_{\text{spec}} = 0.00$-$0.04$ range. This pattern, also observed by \citet{2020MNRAS.491.3767S} in SDSS data, was attributed to the misidentification of stars as galaxies in the SDSS photometric pipeline within this low redshift range.
 Our results show similar discrepancies in the DR9 dataset, suggesting this issue persists.

\section{Conclusion} \label{sec:conclusion}
%\section{Discussion and Conclusions} \label{sec:conclusion}

In this study, we proposed a two-stage Swin Transformer model for effectively identifying ring galaxy candidates in large-scale galaxy image datasets. The model comprises two binary classifiers: the first classifier performs an initial screening of potential ring galaxy candidates, while the second classifier further refines the results to eliminate misclassified samples and improve identification accuracy.
In both classification stages, we experimented with and discussed the impact of sample quantity and the positive-to-negative sample ratio on model performance. 
%To optimize model performance, we experimentally evaluated the ratio of positive to negative samples during training. 

When applying the trained Swin Transformer model to the DR9 of the Legacy Surveys, we focused on 573,668 images with $z\_spec = 0.01$-$0.20$ and $mag_r$ \textless 17.5. After the two-stage classification process, we identified 8052 new ring galaxies with a $Precision$ exceeding 64.87\%.
Our model's performance is comparable to recent machine learning methods, such as the 58.9\% $Precision$ reported by \citet{Krishnakumar_2024} in their search for ring galaxies in \citet{2020ApJS..251...28G} catalog, where 1967 ring galaxies were identified. Notably, our approach does not rely on large-scale simulated data.

In summary, our study introduces an effective method for identifying ring galaxies in large-scale surveys, which is also applicable for locating specific targets in extensive image datasets, particularly with small sample sizes. We have compiled a catalog of 8052 newly discovered ring galaxies, providing solid support for future research and validation efforts. In the future, once sufficient spectral data from DESI becomes available, we will conduct further spectroscopic analysis of the ring galaxies identified to gain a deeper understanding of their physical properties, formation mechanisms, and evolutionary histories.

\section*{Acknowledgments}
This work is supported by the National Natural Science
Foundation of China (NSFC) under grant Nos. 11873037,
U1931209, and 11803016, the Natural Science Foundation of
Shandong Province under ZR2022MA076, ZR2022MA089, ZR2024MA063.
NL acknowledge the support from the science research grants from the China Manned Space Project (No. CMS-CSST-2021-A01), the CAS Project for Young Scientists in Basic Research (No. YSBR-062) and the Ministry of Science and Technology of China (No. 2020SKA0110100).

The Legacy Surveys consist of three individual and complementary projects: the Dark Energy Camera Legacy Survey (DECaLS; Proposal ID \#2014B-0404; PIs: David Schlegel and Arjun Dey), the Beijing-Arizona Sky Survey (BASS; NOAO Prop. ID \#2015A-0801; PIs: Zhou Xu and Xiaohui Fan), and the Mayall z-band Legacy Survey (MzLS; Prop. ID \#2016A-0453; PI: Arjun Dey). DECaLS, BASS and MzLS together include data obtained, respectively, at the Blanco telescope, Cerro Tololo Inter-American Observatory, NSF’s NOIRLab; the Bok telescope, Steward Observatory, University of Arizona; and the Mayall telescope, Kitt Peak National Observatory, NOIRLab. Pipeline processing and analyses of the data were supported by NOIRLab and the Lawrence Berkeley National Laboratory (LBNL). The Legacy Surveys project is honored to be permitted to conduct astronomical research on Iolkam Du’ag (Kitt Peak), a mountain with particular significance to the Tohono O’odham Nation.

NOIRLab is operated by the Association of Universities for Research in Astronomy (AURA) under a cooperative agreement with the National Science Foundation. LBNL is managed by the Regents of the University of California under contract to the U.S. Department of Energy.

This project used data obtained with the Dark Energy Camera (DECam), which was constructed by the Dark Energy Survey (DES) collaboration. Funding for the DES Projects has been provided by the U.S. Department of Energy, the U.S. National Science Foundation, the Ministry of Science and Education of Spain, the Science and Technology Facilities Council of the United Kingdom, the Higher Education Funding Council for England, the National Center for Supercomputing Applications at the University of Illinois at Urbana-Champaign, the Kavli Institute of Cosmological Physics at the University of Chicago, Center for Cosmology and Astro-Particle Physics at the Ohio State University, the Mitchell Institute for Fundamental Physics and Astronomy at Texas A\&M University, Financiadora de Estudos e Projetos, Fundacao Carlos Chagas Filho de Amparo, Financiadora de Estudos e Projetos, Fundacao Carlos Chagas Filho de Amparo a Pesquisa do Estado do Rio de Janeiro, Conselho Nacional de Desenvolvimento Cientifico e Tecnologico and the Ministerio da Ciencia, Tecnologia e Inovacao, the Deutsche Forschungsgemeinschaft and the Collaborating Institutions in the Dark Energy Survey. The Collaborating Institutions are Argonne National Laboratory, the University of California at Santa Cruz, the University of Cambridge, Centro de Investigaciones Energeticas, Medioambientales y Tecnologicas-Madrid, the University of Chicago, University College London, the DES-Brazil Consortium, the University of Edinburgh, the Eidgenossische Technische Hochschule (ETH) Zurich, Fermi National Accelerator Laboratory, the University of Illinois at Urbana-Champaign, the Institut de Ciencies de l’Espai (IEEC/CSIC), the Institut de Fisica d’Altes Energies, Lawrence Berkeley National Laboratory, the Ludwig Maximilians Universitat Munchen and the associated Excellence Cluster Universe, the University of Michigan, NSF’s NOIRLab, the University of Nottingham, the Ohio State University, the University of Pennsylvania, the University of Portsmouth, SLAC National Accelerator Laboratory, Stanford University, the University of Sussex, and Texas A\&M University.

BASS is a key project of the Telescope Access Program (TAP), which has been funded by the National Astronomical Observatories of China, the Chinese Academy of Sciences (the Strategic Priority Research Program “The Emergence of Cosmological Structures” Grant \# XDB09000000), and the Special Fund for Astronomy from the Ministry of Finance. The BASS is also supported by the External Cooperation Program of Chinese Academy of Sciences (Grant \# 114A11KYSB20160057), and Chinese National Natural Science Foundation (Grant \# 12120101003, \# 11433005).

The Legacy Survey team makes use of data products from the Near-Earth Object Wide-field Infrared Survey Explorer (NEOWISE), which is a project of the Jet Propulsion Laboratory/California Institute of Technology. NEOWISE is funded by the National Aeronautics and Space Administration.

The Legacy Surveys imaging of the DESI footprint is supported by the Director, Office of Science, Office of High Energy Physics of the U.S. Department of Energy under Contract No. DE-AC02-05CH1123, by the National Energy Research Scientific Computing Center, a DOE Office of Science User Facility under the same contract; and by the U.S. National Science Foundation, Division of Astronomical Sciences under Contract No. AST-0950945 to NOAO.

%\vspace*{\baselineskip}
%\indent

%\indent

\bibliography{ring-galaxy-sample631}{}
\bibliographystyle{aasjournal}




\end{document}